\documentclass[a4paper]{amsproc}
\usepackage[utf8]{inputenc}
\usepackage{amssymb}
\usepackage{amscd} %% Package for commutative diagrams
\def\fe{\mathsf{e}} %fajlagos energia
\def\fs{\mathsf{s}} %fajlagos entrópia
\def\fv{\mathsf{v}} %fajtréfogat
\def\rhe{\rho_{e}}   %belső energia sűrűség
\def\rs{\rho_{s}}   %entrópiasűrűség
\def\dsigma{\tilde\sigma} %feszültség dinamikus
\def\asigma{\sigma_{an}} %feszültség analeasztikus
\def\vsigma{\sigma_{visc}} %feszültség viszkózus
\def\d{\text{d}} %dinamikus nyomás

\def\re#1{(\ref{#1})}   %% Note: AMSTeX's  \eqref  also does  (\ref{#1})

\newcommand{\pd}[2]{\frac{\partial #1}{\partial #2}}

\theoremstyle{plain}

\theoremstyle{definition}

\theoremstyle{remark}
 
 \numberwithin{equation}{section}

\renewcommand{\geq}{\geqslant}

\title[Thermodynamic gradient elasticity]{Thermodynamically consistent gradient elasticity with an internal variable}

\subjclass[2010]{74A15;74A60}

\keywords{nonequilibrium thermodynamics, generalized continua, gradient elasticity}

\author[Ván]{\bfseries Peter Ván}

\address{
 Wigner Research Centre for Physics, Department of Theoretical Physics, Budapest, Hungary, \\
 Budapest University of Technology and Economics, Faculty of Mechanical Engineering, Department of Energy Engineering, Budapest, Hungary\\
 Montavid Thermodynamic Research Group, Budapest, Hungary
}
\email{van.peter@wigner.hu}

\begin{document}

{\begin{flushleft}\baselineskip9pt\scriptsize THEORETICAL AND
APPLIED MECHANICS\newline %Volume ?? (201?) Issue 1, ***--***\hfill
%DOI:
\end{flushleft}}
\vspace{18mm} \setcounter{page}{1} \thispagestyle{empty}

\begin{abstract}
The role of thermodynamics in continuum mechanics and the derivation of the proper constitutive relations is a discussed subject of Rational Mechanics. The classical literature did not use the accumulated knowledge of thermostatics and was very critical with the heuristic methods of irreversible thermodynamics. In this paper, a  small strain gradient elasticity theory is constructed with memory effects and dissipation. The method is nonequilibrium thermodynamics with internal variables; therefore, the constitutive relations are compatible with thermodynamics by construction. Thermostatic Gibbs relation is introduced for elastic bodies with a single tensorial internal variable. The thermodynamic potentials are first-order weakly nonlocal, and the entropy production is calculated. Then the constitutive functions and the evolution equation of the internal variable is constructed. The second law analysis has shown a contribution of gradient terms to the stress, also without dissipation.  % It is argued that the presented results are in concordance with the principles of Rational Mechanics, in particular with material frame indifference and with the rigorous exploitation of the second law inequality. 
\end{abstract}

%%\dedicatory{ If you want do dedicate your paper to somebody}

\maketitle

\section{Introduction}

Rational mechanics has been started with a sharp criticism toward the mathematics used in continuum mechanics \cite{TruTou60b, TruNol65b}. However, the criticism went far beyond some suggested methodological improvements. A complete reorganisation of fundamental aspects was proposed from two main points of view: the representation of spacetime and the representation of thermodynamic principles. In the following, we survey these aspects and argue that methods of nonequilibrium thermodynamics are under the principles of continuum mechanics in general and with elasticity in particular. In \cite{Van19p} small strain linear elasticity was treated, and the most important aspects were summarised. In the following, after the discussion of objectivity and thermodynamic aspects, we extend these previous results and obtain a dissipative gradient theory of small strain elasticity with a weakly nonlocal internal variable. 

\section{Objectivity}

Objectivity is the concept of spacetime representation of physical quantities and laws. Spacetime representation is particularly important in nonrelativistic continuum mechanics and closely related to the principle of material frame indifference \cite{Fre09a}, a concept whose mathematical formulation was developed by Noll. This formulation requires transformation rules between reference frames \cite{Nol67a}. Later on, his deeper analysis led to a mathematical structure, based on affine spaces, but without a detailed spacetime model \cite{Nol04m, Nol06a, NolSeg10a}. In a complete spacetime formulation the physical quantities, their governing differential equations and also the constitutive functions can be given in an absolute form, that is without any reference frames and independently of the flow of the material \cite{Mat93b}.  In this framework, a physical theory is reference frame independent by construction. However, a complete formulation meets conceptual difficulties, including the simplest possible case of one-component simple fluids. One of the key difficulties is to establish a frame independent concept of energy, because kinetic energy, including the density of kinetic energy of a one-component simple fluid, $K= \frac{\rho v^2}{2}$, cannot be frame independent, being defined by the relative velocity. Therefore it is zero when the reference frame is fixed to the fluid and not zero in any other reference frames. An absolute formulation requires the use of four quantities, spacetime vectors and tensors, also in a nonrelativistic framework. There the time is absolute, but space is relative. Therefore a spatial, three-dimensional space vector or tensor cannot be frame independent, but a four-dimensional spacetime vector or tensor can. This observation also appears indirectly in the transformation rule based definition of Noll, too \cite{MatVan06a}. In a complete spacetime formulation, transformation rules are not parts of the definition of objectivity and can be avoided completely  \cite{Van17a}. Then the frame independence of the Gibbs relation and the entropy production can be proved. This formulation is compatible with the related concepts of Rational Extended Thermodynamics  \cite{MulRug98b,Rug89a,VanEta19a} and also with special relativity \cite{VanBir12a}. The consequences of spacetime formulation for the kinematics lead to a generalisation that does not require the existence of a stress-free, relaxed state of any continua in a finite deformation theory \cite{FulVan12a}. 

It is remarkable that the formalism of four quantities is not always necessary and can be safely avoided. It is also important to understand how far can we go with the help of our customary three-dimensional vectors and tensors. The following aspects of nonrelativistic spacetime are to be considered:
\begin{itemize}
    \item The four vector representation of physical quantities makes inevitable, that the density of an extensive physical quantity, $\rho_{ext}$ and its current density, $\bf{j}_{ext}$, are parts of the same absolute four quantity. This is apparent also in a nonrelativistic, more properly, Galilean relativistic framework. For example conductive and convective current densities, $\bf{j}_{cond}$ and $\rho_{ext} \mathbf{v}$, are related with the formula $\mathbf{j}_{tot} = \mathbf{j}_{cond}+\rho_{ext} \mathbf{v}$. This is the transformation rule between comoving and laboratory frames of the spatial part of a four vector, where the timelike component is the density, and the spacelike component is the current density in a particular reference frame. 
    \item Spacetime derivatives are four covectors. Spacelike components of four covectors are absolute, but timelike components transform, and they differ depending on reference frames. For example the relation of a local, partial time derivative in a laboratory frame, $\partial_t$ to a substantial time derivative, $\frac{d}{dt}$ with the relation $\frac{d}{dt} =\partial_t + \mathbf{v}\cdot\nabla$ is a transformation rule of the timelike component of the spacetime derivative between a laboratory reference frame and a reference frame, comoving with the fluid. Then, the gradients are spacelike covectors and do not transform at all. This fact is well hidden using Noll's definition.
\end{itemize}

A consequence of these observations is, that gradient dependent constitutive functions are frame independent, but one should be careful with time derivatives. It is also easy to comprehend that a balance is a four divergence; therefore, it is frame independent. 

In the following, the application of these simple rules ensures that we obtain frame independent theories with the usual tools of three-dimensional tensor analysis. The simple but not customary four-dimensional affine spaces of nonrelativistic spacetime are not necessary, one can focus on the main subject of the paper, on the formulation of thermodynamic principles.

\section{Second law of thermodynamics}

Rational mechanics considers the second law as a foundation of constitutive modelling in continuum mechanics. However, concepts from thermostatics, like the Gibbs relation with differentials, were abandoned, and the convenient and simple methods of classical irreversible thermodynamics with thermodynamic fluxes and forces are mostly rejected as mathematically inexact \cite{Tru84b}. The criticism was well deserved, obscure concepts cannot lead far, and the development of irreversible thermodynamics slowed down: despite many efforts, classical irreversible thermodynamics was unable to enter into the realm of continuum mechanics. The research of the most critical challenges, time and space nonlocalities that is rheology and gradient theories, do not use irreversible thermodynamics, except some notable but not influential results \cite{Klu62a1, Klu62a2, KluCia78a, CiaKlu79a, Ver97b}. The breakthrough of the last fifty years, Extended Thermodynamics, was the result of ideas from kinetic theory and much less the consequence of a rigorous rational methodology \cite{JouAta92b, LebEta08b}. The rigorous approach encountered the inadequate formulation of objectivity, has chosen kinetic theory as a basis and established a concept of objectivity rejecting the definition of Noll \cite{Mul72a, MulRug98b, Mus12a}.

On the other hand, the rigorous mathematical approach did not result in the expected general and universal theory extending the modelling power of continuum theories but effectively blocked some research directions, in particular toward weakly nonlocal extensions. In their influential paper, Coleman and Gurtin proved that weakly nonlocal internal variables are incompatible with second law \cite{ColGur67a} and in another paper, Gurtin argued that gradient elasticity is incompatible with the second law \cite{Gur65a}. Therefore theories of spatial interactions were developed in a different direction, mostly without the direct constructive application of thermodynamic principles \cite{Min65a, CarEta03a, PapFor06a, For08a, BerFor14a, CorEta16a, For20a, VarAif94a, Aif03a, AskAif11a} or with the help of brand new concepts, like phase fields \cite{PenFif90a, PenFif93a}, interstitial working, extra fluxes \cite{DunSer85a, FabEta11a}  or microforce balance \cite{Gur96a,Gur00b,Gur03a}. In these theories, the importance of the second law varies, but in general, does not play a constructive role. The big idea of the principle-based rational approach has encountered difficulties with the complicated memory functional material models and also encountered mathematical problems, because the constitutive theory leads to unavoidable and physically unacceptable instabilities in gradient elasticity \cite{DunFos74a, Jos81a, Mul07b}. 

However, one may realise that the problem with rational methodology is not the use of mathematics, but with the rigid attitude of finding the correct physical starting points. The conditions of a theorem are to be scrutinised from a physical point of view and modified if necessary. For example, the rigorous methods of second law analysis, the Coleman-Noll and Liu procedures \cite{ColNol63a, Liu72a} can be extended to obtain weakly nonlocal constitutive functions with three conditions:
\begin{itemize}
\item The entropy flux is a constitutive quantity, and it is not always equal heat flux divided by temperature.
\item The spatial derivative of a constraint can be considered as an additional constraint, depending on the structure of the constitutive state space.
\item Thermodynamics fluxes and forces are available concepts to solve the entropy inequality.
\end{itemize}

The first condition, the idea to treat the entropy flux as a constitutive quantity is originated form the work of Müller \cite{Mul67a} and later refined by Verhás and Nyíri \cite{Ver83a,Nyi91a1}. It is also a spacetime question, entropy density and entropy flux are parts of the same objective physical quantity, the entropy four vector. Then one can show that classical irreversible thermodynamics is a first-order weakly nonlocal constitutive theory with balances as constraints and the thermodynamic flux-force system appears naturally \cite{Van03a}. The extended Coleman-Noll and Liu procedures are applicable for checking the second law compatibility of weakly nonlocal continuum theories \cite{Van05a, Cim07a} and can be applied for constructing new ones or unifying independent looking theoretical developments, like internal variables with generalised continua or phase field evolution \cite{Van09a, BerEta11a, VanEta14a, Van18bc, Van19bc}. 

Also, the Gibbs relation of thermostatics is a source of information that should be understood; otherwise, we neglect the related accumulated experience from physics and chemistry. Here the critical aspect is the extension of the concept of extensivity for situations where its original definition is seemingly not applicable, for example to elasticity and gradient effects. These are the subjects of the next sections.

\section{Thermostatics of elasticity}

In the following, we use index notation in a small strain theory, where the strain is denoted by $\varepsilon^{ij}$. Here the indices are spatial and upper-lower index pairs denote summation, e.g. $\varepsilon^{i}_{\,i}$ is the trace of the strain tensor. The indices are abstract, that is they do not denote any coordinates, only the tensorial properties of the spatial physical quantities in the three dimensional vector space of the position. The distinction of upper-lower indices is not always essential, because the three dimensional relative space is endowed by and Euclidean metric, therefore one may identify vectors with covectors. This kind of abstract index notation was introduced by Penrose in relativity theories \cite{Pen04b}, and its use and advantage in nonrelativistic (i.e. Galilean relativistic) spacetime was explained in \cite{Van17a}. 

Deformation or strain cannot be extensive thermodynamic state variables in the classical sense. Deformation and strain are locally defined quantities but homogeneous deformation of a finite volume continuum body depends on the shape of the body, therefore thermodynamic potentials of the complete body do not reflect material properties. In some thermodynamic books the Gibbs relation for elasticity appears as an analogon of the fluid Gibbs relation, but specific volume is substituted by strain \cite{Mau99b, Ver97b, Kui94b}. Other handbooks about continuum mechanics do not consider homogeneous bodies at all \cite{Mau99b}, neither when considering thermodynamic requirements \cite{Tru77b, BedEta95b, GurEta10b}. However, strain is a local concept by definition and it is not necessary to start from quantities of the whole body in a continuum thermodynamic approach. The basic question here is the separation of the local strain from local rotation and the local Riemanian metric, that is responsible for the energy changes in the body \cite{FulVan12a}. Starting from local expressions one can build up extensivity from this direction \cite{BerVan17b}. In the following we will introduce only small strains, where these problems can be solved relatively easily. Then the specific entropy, $\fs$, of elastic bodies is the function of the internal energy, $e$,  and the strain, $\varepsilon^{ij}$, a second order symmetric tensor. Its partial derivatives are:
\begin{equation}
\fs = \fs(\fe,\varepsilon^{ij}), \qquad 
\pd{\fs}{\fe} = \frac{1}{T}, \qquad 
\pd{\fs}{\varepsilon^{ij}} = -\fv\frac{\sigma_{ij}}{T}.
\label{pdefs}\end{equation}
Here \(\sigma_{ij}\) is the thermostatic stress, $\fv$ is the specific volume and $T$ is the temperature. Therefore the Gibbs relation is written as
\begin{equation}
\text{d}\fe = T\text{d}\fs+\fv \sigma_{ij} \text{d}\varepsilon^{ij}.
\label{rGr}\end{equation}

The first order Euler homogeneity is ensured by introducing the chemical potential, $\mu$, as
\begin{equation}
\mu := \fe - T\fs - \fv \sigma_{ij} \varepsilon^{ij}.
\label{rExt}
\end{equation}

Then it is easy to obtain the particular expressions with densities as well:
\begin{equation}
\text{d}\rhe = T\text{d}\rs + \sigma_{ij}\text{d} \varepsilon^{ij} + \left(\mu + \frac{\sigma_{ij} \varepsilon^{ij}}{\rho}\right)\text{d}\rho, 
\label{rGrdens}\end{equation}
\begin{equation}
\rhe = T\rs + \sigma_{ij} \varepsilon^{ij} + \mu \rho.
\label{rExtdens}\end{equation}

Here $\rho = 1/\fv$ is the density, $\rhe = \rho e$ is the internal energy density, $\rs = \rho s$ is the entropy density. It is easy to prove these expressions according to \re{rGr} and \re{rExt}. Density changes are also not negligible in case of small strains. The quantities for the entire elastic body are calculated by multiplying the strain with the mass of the body $M$ and then $M \varepsilon^{ij}$  will be the bulk, body-related extensive thermodynamic state variable if it is interpretable. Chemical potential is rarely introduced explicitly in continuum mechanics because mass exchange usually does not play a role. One can convert the previous expressions with the help of free energy density, $f$, defined by the following Legendre transformation: $f = \rhe- T\rs =   \sigma_{ij} \varepsilon^{ij} + \mu \rho$, and then the Gibbs relation for densities \re{rGrdens} can be written as
\begin{equation}
\text{d}f = - \rs\text{d}T + \sigma_{ij}\text{d} \varepsilon^{ij} + \frac{f}{\rho}\text{d}\rho, 
\qquad \rightarrow \qquad
\rho\text{d}\frac{f}{\rho} = - \rs\text{d}T + \sigma_{ij}\text{d} \varepsilon^{ij}.
\label{frGrdens}\end{equation}
Seemingly there is no need for the chemical potential at all. It is substituted by the specific free energy on the left-hand side of the previous expression. One can avoid the use of entropy and chemical potential starting from free energy in thermodynamical considerations of continuum mechanics \cite{Sil97b, GurEta10b, BerVan17b}. There one should take care of the variables of the previous functions, for example, free energy density has the natural variables: temperature, $T$, strain $\varepsilon^{ij}$ and density $\rho$, as it is apparent from the first Gibbs relation of \re{frGrdens}. With the classical thermodynamical, differential based representation of Gibbs relation one can keep the flexibility of classical thermodynamics when changing the variables. One can see, that with specific quantities the number of variables is reduced, while the free energy density is a function of three variables, $f(T,\varepsilon^{ij},\rho)$, the specific free energy is the function of two, only, $\frac{f}{\rho}(T,\varepsilon^{ij})$. This is the consequence of the extensivity, the first-order Euler homogeneity of the entropy of homogeneous bodies, here considered directly for locally defined specific quantities and densities \cite{BerVan17b}. 

For ideal elastic bodies elastic energy is to be subtracted from the internal energy. Then, one may have two basic choices. In the following, the specific quantities are preferred, therefore specific elastic energy is to be subtracted from the specific internal energy, and the specific entropy is given in the following form 
\begin{equation}
s(e,\varepsilon^{ij}) =  s\left(e- \mu \varepsilon^{ij}\varepsilon_{ij} - \frac{\lambda }{2}(\varepsilon^{i}_{\,i})^2\right),
\label{idels}\end{equation}
where the Lam\'e coefficients are $\hat{\mu} = \rho \mu$ and $\hat{\lambda} = \rho\lambda$, because $\rho_{ela} = \hat\mu \varepsilon^{ij}\varepsilon^{ij} + \frac{\hat\lambda }{2}(\varepsilon^{ii})^2$ is the density of the elastic energy. Then using the definition of temperature as the derivative of the entropy in \re{pdefs}, and assuming constant $\mu, \lambda$ parameters, one obtains, that
\begin{equation}
\sigma^{ij} = 2 \hat\mu \varepsilon^{ij} + \hat\lambda\varepsilon^{k}_{\,k}\delta^{ij}.
\label{sstr}\end{equation}

Here the elastic moduli $\mu$ and $\lambda$ are nonnegative, because of the concavity of the entropy. $\delta^{ij}$ is the Kronecker delta, the identity tensor with abstract index notation. A consequence of this calculation is that the Lam\'e coefficients are proportional to the density. 

We have seen that specific quantities are the most straightforward starting points for constructing thermodynamic potentials in continua. In the following, we further develop this observation.

\section{Thermostatics of gradient elasticity with a gradient internal variable}

The basic problem of using gradients of physical quantities as thermodynamic state variables is a similar shape dependence than in case of deformation. After the previous considerations it is straightforward to introduce the necessary modifications and extend the elastic Gibbs relation. Now the specific entropy depends on the internal  energy, the strain gradient,  $\partial_k\varepsilon^{ij}$, and also on an internal variable $\xi^{ij}$ and its gradient $\partial_k\xi^{ij}$, therefore, $\fs = \fs(\fe,\varepsilon^{ij},\partial_k\varepsilon^{ij},\xi^{ij},\partial_k\xi^{ij})$. Let us denote the partial derivatives of the entropy as
\begin{equation}
\pd{\fs}{\fe} = \frac{1}{T}, \quad 
\pd{\fs}{\varepsilon^{ij}} = -\fv\frac{\sigma_{ij}}{T},\quad 
\pd{\fs}{(\partial_k\varepsilon^{ij})} = \fv\frac{S^k_{\,\,ij}}{T},\quad 
\pd{\fs}{\xi^{ij}} = \fv y_{ij}\quad 
\pd{\fs}{(\partial_k\xi^{ij})} = \fv Y_{\,\,ij}^{k}.
\label{genpdefs}\end{equation}
Here the physical quantities denoted by \(S^k_{\,\,ij},y_{ij},Y_{ij}^{\,\,k}\) are particular intensive thermodynamic state functions in analogy with thermostatic terminology. We will see, that some of the properties of usual intensive quantities are preserved. The corresponding Gibbs relation is written as
\begin{equation}
\text{d}\fe = T\text{d}\fs + \fv \sigma_{ij} \text{d}\varepsilon^{ij} - 
\fv S^k_{\,\,ij} \text{d}\partial_k\varepsilon^{ij}  - 
\fv T y_{ij}\text{d}\xi^{ij} -
\fv T Y_{\,\,ij}^{k} \text{d}\partial_k\xi^{ij}.
\label{genrGr}\end{equation}

This will be our basic formula for the second law inequality and constructing thermodynamic compatible evolution equation for $\xi^{ij}$ and constitutive function for the stress and heat flux. The explicit appearance of temperature in the definition of the internal variable related intensive quantities is not fundamental, it expresses our traditional expectation that the strain directly contributes to the energy, as we have seen in \re{sstr}, but the internal variable may influence the entropy more directly.

\section{Entropy production of gradient elasticity with a weakly nonlocal internal variable}

The entropy inequality is conditional. The fundamental balances are the conditions for the entropy inequality. In our case they are the conservation of mass, the conservation of momentum and the conservation of energy. The conservation of mass, the continuity equation  is written as
\begin{equation}
\label{bal_mass}
\dot \rho + \rho \partial_i v^i = 0,
\end{equation}
where the dot denotes the substantial, comoving time derivative, that is $\dot \rho = \partial_t \rho + v^i\partial_i \rho$. Here $\partial_t$ and $\partial_i$ are the partial time derivative and the gradient, respectively. $v^i$ is the velocity field of the continuum, defined in the usual way as mass and momentum flow \cite{VanEta17a}. The balance of momentum is 
\begin{equation}
\label{bal_mom}
\rho\dot v^i - \partial_j\dsigma^{ij} = 0^i.
\end{equation}

Here $\dsigma^{ij}$ is the stress tensor. The conservation of the moment of momentum is assumed; therefore the stress is symmetric, $\dsigma^{ij} = \dsigma^{ji}$. Note, that the stress in the momentum balance can be different of the static stress $\sigma^{ij}$, given as the derivative of the entropy function in \re{pdefs} and for ideal elasticity in particular in \re{sstr}. The balance of internal energy follows as (see, e.g. \cite{Gya70b})
\begin{equation}
\label{bal_inte}
\rho\dot e + \partial_i q^{i} = \dsigma^{ij}\partial_iv_j,
\end{equation}
where $q^i$ is the heat flux, the conductive current density of the internal energy. It is different from the energy flux $\hat q^i = q^i -\dsigma^{ij}v_j$, the conductive current density of the total energy. Here the antisymmetric part of the velocity gradient tensor, which is the curl of the velocity field, does not play a role, because of the symmetry of the stress. 

A fourth condition that must be considered is the small strain version of the compatibility condition, that is 
\begin{equation}
\dot \varepsilon^{ij} = \frac{1}{2}(\partial^{i}v^j+\partial^jv^i),
\label{compcond}\end{equation}
The substantial time derivative of the strain is the symmetric part of the  velocity gradient. 

The last condition is less evident, and it is the evolution equation of the internal variable, expressed explicitly as  
\begin{equation}
\dot \xi^{ij} = f^{ij}(\fe,\varepsilon^{ij},\partial_k\varepsilon^{ij},\xi^{ij},\partial_k\xi^{ij}).
\label{ev_intv}\end{equation}
This condition expresses that the evolution equation of the internal variable is unknown and is to be determined constitutively, with the help of the second law. The restriction from the second law is universal, independent of the particular material structure which defines the internal variable, and this possibility is the most important consequence of recent thermodynamic investigations.

The entropy balance expresses the second law in the form of the following conditional inequality
\begin{equation}
\label{bal_entr}
\rho\dot s + \partial_i J^{i} = \Sigma \geq 0,
\end{equation}
where the above conditions, \re{bal_mass}-\re{compcond} are to be considered with the simplest possible way, by direct substitution. Then the calculation of the entropy production is straightforward if the entropy flux is identified. For that purpose, the classical method of irrversible thermodynamics is applied \cite{GroMaz62b}. The generalisation to the weakly nonlocal, gradient dependent case is straightforward, see, e.g. \cite{BerEta11a, Van18bc}. The direct application of the Gibbs relation \re{genrGr} requires to use only the balance of the internal energy, \re{bal_inte} and the compatibility condition \re{compcond}. The balance of momentum is considered through the internal energy, and the continuity equation is not necessary because of the use of substantial derivatives and conductive fluxes. This simplification, common in fluid mechanics, considers comoving quantities, separating the changes in the various fields due to the motion of the continuum from the changes of material origin. The whole procedure and also the entropy production is absolute, reference frame and flow-frame independent. A more detailed explanation of the related objectivity issues was given in the introduction and also \cite{Van17a}. The extension of those calculations for the present case is straightforward. 
\begin{gather}
\rho\dot s(e,\varepsilon^{ij},\partial_k\varepsilon^{ij},\xi^{ij},\partial_k\xi^{ij}) = 
    \frac{\rho\dot e}{T} - \frac{\rho}{\rho T} {\sigma}_{ij} \dot{\varepsilon}^{ij} + 
    \frac{S^{k}_{\,\,ij}}{T}  ({\partial_k\varepsilon}^{ij}\dot) + 
    {y_{ij}}\dot{\xi}^{ij} + 
    Y^k_{\,\,ij}({\partial_k\xi}^{ij}\dot) = \nonumber\\
= -\partial_k \left(\frac{q^k - S^{k}_{\,\,ij}\dot\varepsilon^{ij}}{T} - Y^{k}_{\,\,ij}\dot\xi^{ij}\right) +     
    \partial_k\left(\frac{1}{T}\right)(q^k - S^{k}_{\,\,ij}\dot\varepsilon^{ij}) -\nonumber\\
    \frac{\dot{\varepsilon_{ij}}}{T} \left(\sigma^{ij} +     \partial_k S^{kij}\right) +
    \frac{\partial_j v_i}{T} \left(\dsigma^{ij} -
        S_{\,\,kl}^{j}\partial^{i}\varepsilon^{kl}-
        TY_{\,\,kl}^{j} \partial^{i}\xi^{kl}\right) +\nonumber\\
    f^{ij}\left(y_{ij} - \partial_k Y^{\,\,k}_{ij}\right)\geq 0.
\end{gather}

Here we can identify the entropy flux as $J^k = \frac{q^k - S^{k}_{\,\,ij}\dot\varepsilon^{ij} - TY^{k}_{\,\,ij}\dot\xi^{ij}}{T}$, and a modified heat flux with an extra term $\hat q^k = q^k - S^{k}_{\,\,ij}\dot\varepsilon^{ij}$. Introducing the compatibility condition, \re{compcond}, one obtains for  the entropy balance:
\begin{gather}
\rho \dot s + 
    \partial_k \left(\frac{q^k - S^{k}_{\,\,ij}\dot\varepsilon^{ij} - 
    T Y^{k}_{\,\,ij}\dot\xi^{ij}}{T}\right) = \nonumber\\
\left(q^k- S^{k}_{\,\,ij}\dot\varepsilon^{ij}\right)\partial_k\left(\frac{1}{T}\right)  +\nonumber\\
    \frac{\dot{\varepsilon_{ij}}}{T} \left(\dsigma^{(ij)} - \sigma^{ij} -
    \partial_k S^{kij} -
    S_{kl}^{\,\,( i}\partial^{j)}\varepsilon^{kl} -
    Y_{kl}^{\,\,( i} \partial^{j)}\xi^{kl}\right) +\nonumber\\
\frac{\partial_{[ j} v_{i ]}}{T} \left(\dsigma^{[ij]} - S_{kl}^{\,\,[j}\partial^{i]}\varepsilon^{kl} -
        TY_{kl}^{\,\,[j} \partial^{i]}\xi^{kl}\right)+
f^{ij}\left(y_{ij} - \partial_kY^k_{\,\,ij}\right)\geq 0.
\end{gather}
Here $( .. )$ denotes the symmetric part of the corresponding tensorial components and $[..]$ denotes the antisymmetric one. In the calculation we used, that the substantial and spatial derivatives do not commute, and the following identity was applied
\begin{equation}
(\partial_k\xi^{ij}\dot) = \partial_k\dot\xi^{ij} - \partial_kv^l\partial_l\xi^{ij}.
\end{equation} 

Here $\asigma^{(ij)} = \dsigma^{(ij)}- \sigma^{ij}- \partial_k S^{kij} - S_{kl}^{\,\,( i}\partial^{j)}\varepsilon^{kl} -   Y_{kl}^{\,\,( i} \partial^{j)}\xi^{kl}$ is the symmetric {\em anelastic stress}. This expression is the extension of the usual viscous stress,  $\vsigma^{ij} = \dsigma^{ij}- \sigma^{ij}$. The additional anelastic terms are  due to the internal variable and also due to the gradient of the strain.  We can see that weak nonlocality leads to couple-stresses both from the strain and internal variable gradients. We will call $\asigma^{[ij]} = \dsigma^{[ij]} - S_{kl}^{\,\,[j}\partial^{i]}\varepsilon^{kl} - TY_{kl}^{\,\,[j} \partial^{i]}\xi^{kl}$ as {\em anelastic couple-stress}.

The heat flux, $\hat q^i$, the dynamic stress, $\dsigma^{ij}$ and the evolution equation of the internal variable, $f^{ij}$, are material dependent constitutive quantities. The entropy inequality determines their functional form. The simplest solution is to assume that $\hat q^i$, $\asigma^{ij}$ and $f^{ij}$ are linear functions of their multipliers in the entropy inequality, that is we can introduce thermodynamic fluxes and forces, as it is shown in Table 1. 

\begin{table}
\centering
\begin{tabular}{c|c|c|c|c}
       &Thermal  & Mechanical & Couple & Internal \\ \hline
Fluxes & $\hat q^i$ & 
         $\asigma^{(ij)}$ & 
         $\asigma^{[ij]}$ &
          $f^{ij}$ \\ \hline
Forces & $\partial_i\left(\frac{1}{T}\right)$ & 
         $\frac{\dot\varepsilon^{ij}}{T}$ & 
         $\frac{\omega_{[ ji ]}}{T}=\frac{\partial_{[ j} v_{i ]}}{T}$ &
          $y_{ij} - \partial_k Y^k_{\,\,ij}$\\ 
\end{tabular}\\
\caption{Thermodynamic fluxes and forces of weakly nonlocal anelastic solids.}
\label{ff}
\end{table}

The identification of thermodynamic fluxes and forces must be based on their mathematical properties. Originally, for simple materials, thermodynamic forces have a gradient form, and the thermodynamic fluxes are related to conductive current densities, -- called fluxes -- of the balance form constraints \cite{Van03a}. In general thermodynamic fluxes are to be related to the constitutive functions, while thermodynamic forces are given operators, functions on the constitutive state space. Therefore the mechanical stress is not a force in a thermodynamical sense; it is a thermodynamic flux, related to momentum transport in the material. From a physical point of view thermodynamic fluxes are better considered characterising the deviation from local equilibrium and thermodynamic forces are to be considered as generalisations of gradients, their particular form is influenced by the various constraints and by the structure of the state space. For example, in our case, the thermodynamic force for the internal interaction, related to the internal variable $\xi^{ij}$ is a complete partial functional derivative of the entropy density by the internal variable:
\begin{equation}
y_{ij} - \partial_kY^k_{\,\,ij} =\rho \frac{\partial s}{\partial \xi^{ij}} - \partial_k \left(\rho\frac{\partial s}{\partial(\partial_k \xi^{ij})}\right) = \frac{\delta (\rho s)}{\delta \xi^{ij}}(\fe,\varepsilon^{ij},\partial_k\varepsilon^{ij},\xi^{ij},\partial_k\xi^{ij}).
\end{equation}

Assuming that  the constitutive functions are smooth and isotropic, the general solution of the entropy inequality follows from Lagrange mean value theorem in the following form
\begin{eqnarray}
\hat q^i &=& \Lambda \partial_i \frac{1}{T},\label{Fou}\\
\asigma^{\langle ij \rangle} &=& l_{11} \dot\varepsilon^{\langle ij \rangle}  + l_{12} \left(y_{\langle ij \rangle} - \partial_k Y^k_{\,\,\langle ij \rangle}\right)\nonumber\\
\dot \xi^{\langle ij \rangle} &=& l_{21} \dot\varepsilon^{\langle ij \rangle}   + l_{22} \left(y_{\langle ij \rangle} - \partial_k Y^k_{\,\,\langle ij \rangle}\right)\label{devi}\\
(\asigma)^k_k &=& k_{11} \dot\varepsilon^k_k  + k_{12} \left(y^k_k - \partial_l Y^{lk}_k\right)\nonumber\\
\dot \xi^k_k &=& k_{21} \dot\varepsilon^k_k  + k_{22} \left(y^k_k - \partial_l Y^{lk}_k\right).
\label{spheric}\\
\asigma^{[ ij ]} &=& m_{11} \omega^{ij}   + m_{12} \left(y_{[ ij]} - \partial_k Y^k_{\,\,[ ij ]}\right)\nonumber\\
\dot \xi^{[ ij ]} &=& m_{21} \omega^{ij}   + m_{22} \left(y_{[ ij]} - \partial_k Y^k_{\,\,[ ij ]}\right)\label{forg}\end{eqnarray}
Here the representation theorems of isotropic functions were applied that is the second order spatial tensors were divided into a traceless symmetric, that is deviatoric; trace, that is spherical, and the antisymmetric parts as it is customary in isotropic elasticity. Some of the material coefficients are well known. $\Lambda_F = \Lambda/T^2$ is the Fourier heat conductivity coefficient,, $l_{11}$ and $k_{11}$ are the linear viscoelastic coefficients of a Kelvin-Voigt body. The second law, nonnegativity of the entropy production requires the positive definiteness of the symmetric parts of the coefficient matrices and therefore the following sign restrictions follow:
\begin{gather}
\Lambda, \, l_{11}, \, l_{22}, \, k_{11}, \, k_{22}, \, m_{11}, \, m_{22}\geq 0, \\
\quad l_{11}l_{22}- \frac{l_{12}+l_{21}}{4}\geq 0, \quad 
k_{11}k_{22}- \frac{k_{12}+k_{21}}{4}\geq 0, \quad 
m_{11}m_{22}- \frac{m_{12}+m_{21}}{4}\geq 0.
\label{secsigns}\end{gather}
\re{devi} and \re{spheric} are the weakly nonlocal generalization of the Kluitenberg-Verhás body \cite{AssEta15a}, the fundamental building block a thermodynamic rheology. The difference is that here the thermodynamic forces and fluxes are gradient dependent. It is important to remark, that the symmetry of the Onsagerian coefficient matrices cannot required, therefore $l_{12}\neq l_{21}$, $k_{12}\neq k_{21}$, $m_{12}\neq m_{21}$ in general, as it is experimentally observed in case of rock materials for the deviatoric and spherical parts \cite{MatTak93a,Mats08a,LinEta10p}. The coefficients are not necessarily constant, they may be state dependent, and in general, in a fully nonlinear case, they may depend on the thermodynamic forces, too.

\subsection{Ideal anelastic materials}

It is worth to inspect an important special case of our continuum model, when the material is not dissipative. There are several distinct possibilities. Let us assume now, that internal thermodynamic force, $ y_{ij} - \partial_k Y^k_{\,\,ij}$ is zero and also the heat flux and the anelestic stress is zero.  Then the following constitutive functions and field equations are to be considered:
\begin{eqnarray}
q^k             &=& S^{k}_{\,\,ij}\dot\varepsilon^{ij}, \label{idq}\\
\dsigma^{ij}    &=& \sigma^{ij}+ \partial_k S^{kij} + 
        S_{kl}^{\,\, i}\partial^{j}\varepsilon^{kl} +  
        T Y_{kl}^{\,\, i} \partial^{j}\xi^{kl} \\
0               &=&\rho \frac{\partial s}{\partial \xi^{ij}} - \partial_k \left(\rho\frac{\partial s}{\partial(\partial_k \xi^{ij})} \right) = \frac{\delta (\rho s)}{\delta \xi^{ij}}.
\label{statint}\end{eqnarray} 
In this case  the continuum is not necessarily at rest, the rate of the strain is not necessarily zero, as one can see from the constitutive equations \re{devi}-\re{forg} above. 

A remarkable consequence that the heat flux is not zero substituting \re{idq} into the balance of internal energy one can see the propagation of  internal energy connected to the strain changes, that is
\begin{equation}
\rho\dot e + \partial_k(S^{k}_{\,\,ij}\dot\varepsilon^{ij}) = 0.
\end{equation} 

The internal energy is conserved, because the mechanical power is zero. However, the momentum balance 
has the following form:
\begin{equation}
\rho \dot v^i - \partial_j(\sigma^{ij}+ \partial_k S^{kij} +  S_{kl}^{\,\, i}\partial^{j}\varepsilon^{kl} +      T Y_{kl}^{\,\, i} \partial^{j}\xi^{kl}) = 0
\label{mom_tr1}\end{equation}
However, considering \re{statint} the internal variable related stress term can be converted to force density and the equation transforms to
\begin{equation}
\rho \dot v^i - \partial_j(\sigma^{ij}+ \partial_k S^{kij} +  S_{kl}^{\,\, i}\partial^{j}\varepsilon^{kl}) = -\rho \nabla_\xi^i s
\label{mom_tr2}\end{equation}

Here $\nabla_\xi s = \partial_{\xi^{kl}}s\partial^i\xi^{kl} + \partial_{(\partial_j\xi^{kl})} s\partial^i_j\xi^{kl} $, the partial gradient of the specific entropy, due to the internal variable. If the entropy is additively decomposed into strain and internal variable dependent parts, that is 
$s(\fe,\varepsilon^{ij},\partial_k\varepsilon^{ij},\xi^{ij},\partial_k\xi^{ij}) = s_{ela}(\fe,\varepsilon^{ij},\partial_k\varepsilon^{ij})+s_{int}(\xi^{ij},\partial_k\xi^{ij})$, then the gradient of the second part, $s_{int}$ is a force density, because $\nabla_\xi^i s = \partial_i s_{int}$.

\section{Concluding remarks: internal variables, phase fields and gravitation}

Finally, we have some general remarks.
\begin{itemize}
\item It is remarkable that according to the continuity equation and the compatibility condition, \re{compcond} and \re{bal_mass}, the density and the strain are not independent, because: $\frac{\dot \rho}{\rho} = \dot \varepsilon^i_i$. Therefore $\varepsilon^i_i = \ln \frac{\rho}{\rho_0}$, where $\rho_0$ is constant. However, thermostatics, with the definition of the state variables and thermodynamic potentials precedes the calculation of the entropy production, and the continuity equation is a constraint there.
\item The results of these heuristic calculations can be obtained with more rigorous methods, too. Regarding objectivity see \cite{Ful15m}. Regarding the heuristic methodology of Classical Irreversible Thermodynamics behind the previous calculations, see \cite{Van03a}. It was shown there, that our framework -- first-order weakly nonlocal constitutive state space with balance constraints -- leads to the classical form of the entropy production.
\item Sometimes it is assumed that a gradient itself can be an internal variable see, e.g. \cite{BerEta11a, VanFul12a}. Here it is shown that it can be misleading because surface and bulk contributions differ, and evolution equation and boundary conditions are affected.  
%general Aifantis  very poor, gradient cannot be internal variable: surface contributions, internal power,
\item In a finite deformation framework natural objective derivatives, as Lie derivatives, appear in the evolution of the internal variable (see, e.g. \cite{Van08a}). Also spatial interactions are influenced \cite{BerVan17b}. Here our treatment did not introduce material manifolds. That should be a logical next step in this research.
\end{itemize}

The most remarkable aspect of our treatment is the stress force relation of the ideal solid, expressed in \re{mom_tr1} and \re{mom_tr2}. According to the constitutive relations \re{devi}-\re{forg} and in particular the sign restrictions of \re{secsigns} the evolution of the internal variable is relaxational and eventually becomes static. The developed static structure satisfies \re{statint}. This behaviour is general, and it is independent of the tensorial properties of the internal variable. Then the internal variable related stresses are bulk, and all stress contributions may appear as force density in the momentum balance. This is a natural, {\em dynamic homogenization} process. 

We have mentioned, that without the gradient contributions and the couple-stresses, our treatment is reduced to the  Kluitenberg-Verhás rheological body \cite{AssEta15a, VanEta19a, SzuFul19a, FulEta20a}.  However, the previous considerations open the possibility to generalize the constitutive framework of continuum mechanics into various directions together with the constructive approach of nonequilibrium thermodynamics. Weakly nonlocal extension of the classical state variables, memory effects with internal variables and also the combination of memory effects with gradient effects offer a rich framework of material modelling. For example, the previously mentioned stress-force relation for static internal variables shows the natural connection of microforce balance based material models \cite{FriGur93a, Gur96a, Gur00b}, and theories with internal variables. 
The extension considering spatial nonlocalities of the internal variables in a higher-order weakly nonlocal state spaces leads to phase field theories \cite{Van18bc, Van19bc}. Introducing a second tensorial internal variable leads to dual internal variables and results in the micromorphic theory \cite{BerEta11a, VanEta14a}. 

The generalization of the Fourier law, the heat conduction theory is also straightforward. It leads to experimentally confirmed effects in complex materials at room temperature \cite{BotEta16a, VanEta17a}. The connection between mechanical and thermal effects in this framework of nonequilibrium thermodynamics is a source of experimental and technological predictions \cite{JozsKov19b}.

Finally, I would like to mention one of the most striking consequences of the internal variable approach, highlighting the importance of the fine details of weak nonlocality and the requirement of extensivity. Let us assume, that a part of the internal energy is weakly nonlocal with a square gradient weak nonlocality with the following Gibbs relation:
$$
\d u = T\d s - p\d v = \d e - \d \varphi - \d\left( \frac{\partial^i\varphi\partial_i \varphi}{8\pi G\rho}
\right).
$$
With the previous methods of nonequilibrium thermodynamics, it is easy to show, that $\phi$ is the Newtonian gravitational potential and we obtain a dissipative theory of gravity, which in the ideal case reduces to the usual one, where the field equation for the gravitational potential  $\phi$ is the Poisson equation: $\partial^{i}_i\phi = 4\phi G \rho$. However, the general nondissipative dynamics is more general and also reproduces the field equations of Modified Newtonian Dynamics (MOND) \cite{VanAbe19m1}.

\vspace{1cm}
\noindent
\subsection*{Acknowledgments} {
The work was supported by the grants National Research, Development and Innovation Office - NKFIH 124366(124508), 123815, TUDFO/ 51757/2019-ITM (Thematic Excellence Program) and FIEK-16-1-2016-0007. The research reported in this paper was supported by the Higher Education Excellence Program of the Ministry of Human Capacities in the frame of Nanotechnology research area of Budapest University of Technology and Economics (BME FIKP-NANO). The author thanks Robert Kovács for valuable discussions.}  

\vspace{1cm}

\bibliographystyle{amsplain}%\bibliographystyle{unsrt}
%\bibliography{termo,qm,MatolcsiT,nonloc,VanP,stmech,misc,fracmech}
\providecommand{\bysame}{\leavevmode\hbox to3em{\hrulefill}\thinspace}
\providecommand{\MR}{\relax\ifhmode\unskip\space\fi MR }
\providecommand{\MRhref}[2]{%
  \href{http://www.ams.org/mathscinet-getitem?mr=#1}{#2}
}
\providecommand{\href}[2]{#2}

\end{document}